\documentclass[aps,twocolumn,prl]{revtex4}

\usepackage{graphicx}
\usepackage{bm,color}

\usepackage{graphicx}
\usepackage{amsmath}
\usepackage{bm,color}
\usepackage{multirow}

\newcommand{\be}{\begin{eqnarray}}
\newcommand{\ee}{\end{eqnarray}}

\begin{document}

\title{Interaction Induced Topological Charge Pump}
\date{\today}

\author{Yoshihito Kuno}
\author{Yasuhiro Hatsugai}
\affiliation{Department of Physics, University of Tsukuba, Tsukuba, Ibaraki 305-8571, Japan}

\begin{abstract}
Based on a topological transition of the symmetry protected topological phase (SPT),
an interaction induced topological charge pump (iTCP) is proposed
with the symmetry breaking parameter as a synthetic dimension.
It implies that the phase boundary of the SPT is the topological obstruction
although iTCP and the gap closing singularity is stable for symmetry breaking perturbations.
We have confirmed the bulk-edge correspondence for this iTCP using DMRG
for the  Rice-Mele model with nearest-neighbor interactions.
As for a realization in optical lattices, an interaction sweeping pump protocol is proposed as well.
\end{abstract}


\maketitle
\textit{Introduction.---}
Topological charge pump (TCP) \cite{Thouless} is one of the fundamental 
topological phenomena, which includes essence of the topological condensed matter \cite{TKNN,Hatsugai_BEC}. 
Since the TCP was first proposed by Thouless about four decades ago \cite{Thouless}, 
it has been rarely verified experimentally. 
However recent experimental systems: coldatoms in an optical lattice \cite{Lohse,Nakajima,Schweizer} and  photonic crystals \cite{Kraus} have enabled to realize the TCP.
The experimental realizations have made the study of TCP as one of the most active current topic. 
In particular, motivated by the significant controllability in recent coldatom and photonic systems\cite{Ozawa,Cooper}, 
TCP has been focused theoretically from a new point of view. 
Although roles of the edge states have never been discussed as for TCP, 
the bulk--edge correspondence (BEC) of TCP of the non-interacting fermions
was reconsidered 
in \cite{Hatsugai}.
Unlike with various topological phenomena where bulk topological number
is hidden and the edge states are
physical observables, physical observables of TCP is a bulk
current and the edge states are hidden (never pumped in a finite speed pump).
We here firstly establish BEC of TCP for the interacting system after a general proposal of TCP based on the SPT phase transition. 

So far, motivated by recent experimental successes of TCP, various theoretical works of TCP have been reported. 
TCP with Hubbard interaction \cite{Nakagawa}, interacting bosonic systems \cite{RLi,YKe,Kuno2,Hayward,Greschner} 
and magnon pump \cite{Mei} have been discussed. 
Further randomness and non-adiabaticity has been studied in \cite{Kuno4,Ippoliti, Privitera,Zhou,Wang}.

\begin{figure*}[t]
\centering
\includegraphics[width=18cm]{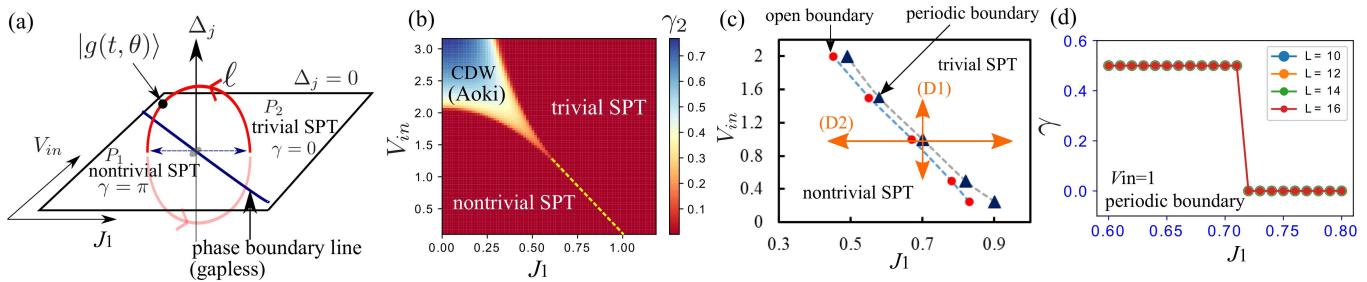}
\caption{(a) A schematic pump protocol oriented from the SPT phase and trivial phase.  
The red loop is a pump protocol loop $\ell$. 
The gapless phase boundary line (topological obstruction) is passing through inside the protocol
loop $\ell$. 
$|g(t,\theta)\rangle$ is a ground-state on the protocol loop. 
(b) MF phase diagram for the interacting SSH model. We set $J_2=1$.
(c) The phase diagram for the interacting SSH model is given by the entanglement spectrum of the open system and the Berry phase of the periodic system. The phase boundary of the open boundary condition is obtained by the entanglement spectrum by the the DMRG with $L=64$ system. The case of the periodic boundary condition is obtained by the Berry phase using ED. 
(d) The Berry phase $\gamma$ in $L/2$-particle system. 
The sharp transition is due to the gap closing by twisting the boundary condition.}
\label{Fig1}
\end{figure*}

In this Letter, we propose interaction induced topological charge pump (iTCP) based on the general scheme. 
Since TCP is independent of any symmetry protection,
existence of nontrivial pump is not trivial a priori.
Then mapping from two-dimensional topologically nontrivial system such as quantum Hall states by replacing one of physical dimensions as a time is useful \cite{Thouless, Hatsugai1}. We here propose another general scheme to realize nontrivial TCP.
Let us start from a one dimensional gapped SPT phase with short range topological order
\cite{Hatsugai1,Pollmann,Wen,Bermudez} that is well characterized by the
symmetry protected Berry phase $i\gamma=\int A $
where $A = \psi ^\dagger d\psi$ and $\psi$ is the ground state \cite{Hatsugai1,EPL-YHIM,PRL-TK-TM-YH}.
The Berry connection $A$ is defined for a twisted boundary condition
$S^1_\theta =\{e^{i\theta}|\theta \in(0,2\pi]\}$.
One may consider this gapped SPT phase
is associated with the twist parameter space $S^1$ which is small in a sense that
physical observables such as the energy are independent of $\theta$ when the system size is infinite \cite{Niu1985,Kudo}.
Further nontrivial $\gamma$ reflecting nontrivial short range order implies existence of edge states when the system has a boundary \cite{Ryu,Hatsugai_Solid}. 
We hereby assume that the iTCP passes through two different SPT phases $P_1$ and $P_2$ 
characterized by different $\gamma$'s implying the number of edge states are different \cite{SPT_dis}. 
Since the quantized Berry phase $\gamma$ is a topological invariant, the energy gap of the system vanishes along any path connecting between $P_1$ and $P_2$. This vanishing point forms a line when the parameter space of the SPT is larger than one (See Fig.~\ref{Fig1} (a)). 
Possible exception can be existence of symmetry breaking phases like charge density wave (CDW) (Aoki phase in the context of Gross-Neuvo model) between $P_1$ and $P_2$. 
The pumping protocol is specified by a loop in a parameter space of the SPT and an extra (synthetic) dimension of the symmetry breaking parameter.
See Fig.~\ref{Fig1} (a). Similar situations are discussed for
the TCP in an extended Bose-Hubbard model \cite{Berg,Rossini}. The gapless line in the parameter space is a topological obstruction of the pumping. It implies that once the nontrivial iTCP is realized, symmetry protection to realize the SPT phase  can be relaxed as far as the gap along the iTCP loop remains open. 

In the following, we shall demonstrate validity of
the proposal by considering a simple Rice-Mele model with inner-unit cell interactions, which is much close to the recent experimental systems \cite{Lohse,Nakajima}.
We shall numerically demonstrate the iTCP and also confirm that the BEC is established for the iTCP.
\textit{Model.---}
The model considered in this work is a generalized Rice-Mele model with nearest-neighbor (NN) interactions:
\begin{eqnarray}
H_{\rm gRM}=\sum^{L-1}_{j=0}&\biggl[& -J_{j}(c^{\dagger}_{j+1}c_{j}+\mbox{h.c.})+\Delta_{j}c^{\dagger}_{j}c_{j}\nonumber\\
&+&V_j \biggl(n_{j}-\frac{1}{2}\biggr)\biggl(n_{j+1}-\frac{1}{2}\biggr)\biggr],\
\label{RM}
\end{eqnarray}
where $c^{(\dagger)}_j$ is a fermion annihilation (creation) operator, $n_{j}=c^{\dagger}_{j}c_j$, $J_j$, $\Delta_j$, and $V_j$ terms are hopping amplitude, on-site potential and NN interactions, respectively. $L$ is a system size.
Here, the parameters $J_{j}$, $\Delta_{j}$ and $V_j$ takes different values for whether $j$ is even or odd. 
For $V_j=0$ case, if we changes $J_{j}$ and $\Delta_j$ dynamically, the above model can be reduced to the the Rice-Mele model \cite{Rice}, which is a standard model of TCP \cite{Asboth}. 
Hereafter we focus only inner-unit cell interactions: $V_{j\in even}=V_{in}$ and $V_{j\in odd}=0$.

\textit{Emergence of iTCP.---}
Let us set $J_{j\in even(odd)}=J_{1(2)}$, where $J_{1(2)}$ is a real value and $\Delta_j=0$, $H_{\rm gRM}$ reduces to the Su-Schrieffer-Heeger (SSH) model \cite{SSH} with the interactions $V_{in}$, which is a lattice analogue of the 
Gross-Neveu model \cite{Aoki,Bermudez,Kuno}. Although Aoki phase as a symmetry breaking phase may exist, it is still a topological obstruction for the iTCP (See Fig.~\ref{Fig1} (b)). 
When $\Delta _i=0$, due to the particle-hole (PH) or 
bond-centered inversion (BCI) symmetries \cite{BCS}, 
a gapped SPT is realized for half-filled case,
which we use it for the iTCP. That is, $\Delta _i$ is a symmetry breaking parameter.
In what follows, we set $J_2=1$.

The phase structure of this model is the starting point to realize the iTCP.
For the SSH model with half-filled case, a mean field (MF) phase diagram is shown in Fig.~\ref{Fig1} (b). 
The MF theory is briefly explained in \cite{SM}. 
The phase diagram has three phases: the nontrivial SPT, trivial SPT and CDW (Aoki) phases. The CDW order corresponds to the Aoki phase in the high-energy physics context \cite{Bermudez,Kuno,Araki}. 
On $J_1$-$V_{in}$ plane where the PH and BCI symmetries are preserved, the gapless phase boundary line necessarily exists between the nontrivial and trivial SPT phases. We do not focus on the CDW order in this Letter. 

For a system without boundary, the Berry phase $\gamma$ is quantized into $Z_2$, 
the nontrivial bulk is characterized by $\gamma =\pi$ \cite{Hatsugai1,Hatsugai2,Guo}.
With an open boundary condition, 
edge modes appear for $\gamma =\pi$ according to the BEC for the one dimensional system. 
By using exact diagonalization (ED), we also confirmed the phase boundary line separating the nontrivial and trivial SPT phases on $J_1$--$V_{in}$ plane, as shown in Fig.~\ref{Fig1} (c). Figure \ref{Fig1} (d) is a typical transition behavior of $\gamma$, where a clear topological phase transition point is determined by the $Z_2$ Berry phase $\gamma$ without any significant system size dependence.  
This indicates that the gapless phase boundary line exists as shown in Fig.~\ref{Fig1} (b) and (c) \cite{SPT_transition}.
The SPT phase boundary on $J_1$--$V_{in}$ parameter space also deviates from $J_1=1$ line (the transition point at the non-interacting case) \cite{SPT_shift}.
It implies various possibility of the iTCP protocols. 

Using the gapless phase boundary line on $J_1$--$V_{in}$ plane as a topological obstruction, various pump protocols to exhibit the TCP can be considered by the loop $\ell$ in $J_1$--$V_{in}$--$\Delta_{j}$ space (See Fig.~\ref{Fig1} (a)). 
The pump protocol specified by the loop $\ell$ is parameterized by  time $t$. 
Together with the small dimension $S^1$ of the twist $\theta$ and  the loop $\ell$ the ground state $|g(\theta,t)\rangle$
is defined on a torus $T^2=S^1\times\ell$.

For concreteness, we set $\Delta_j(t)\equiv (-1)^{j}\Delta_{0} \sin (2\pi t/T)$, where $T$ is a period of the pump. 
Together with $\Delta_j(t)$, we can constitute a pump protocol loop by dynamically varying the interaction $V_{in}$ and/or a hopping ratio $J_{1}/J_{2}$. The concrete form will be given later. 
As for an experimental realization, pumping speed needs to be slow enough compared with the bulk gap. We need this to guarantee the pump to be adiabatic. Note that the appearance of the edge states implies that the system is gapless. Then the system with boundaries cannot be adiabatic in a realistic pump. The edge states are useful for a theoretical understanding of TCP, but never observed directly. The pump loop connects the nontrivial and trivial SPT phases without crossing the gapless phase boundary line. 

The topological invariant of the iTCP is given by the Chern number in a temporal gauge
as
\begin{eqnarray}
i{\bar A}^{(t)}_{\theta}(T)=C,\nonumber 
\end{eqnarray}
where ${\bar A}^{(t)}_{\theta}(t)\equiv \frac{1}{2\pi}\int^{\pi}_{-\pi}A^{(t)}_{\theta}(\theta,t)d\theta$ is a $\theta $ averaged Berry connection in the temoporal gauge 
$A^{(t)}_{\theta}(\theta,t)$ (specified uniquely by $A^{(t)}_t =0$) as \cite{Hatsugai}
\begin{eqnarray}
A^{(t)}_{\theta}(\theta,t)=A_{\theta}(\theta,t)-\partial_{\theta}\int^{t}_{0}A_{t}(\theta,\tau)d\tau-A_{\theta}(\theta,0)\nonumber.
\label{temporal}
\end{eqnarray}
where $A_{t}(\theta,\tau)$ and $A_{\theta}(\theta,\tau)$ are  Berry connections in arbitrary gauge. Note that the Berry connection in the temporal gauge is not periodic in time even though $A_{t(\theta)}(\theta,t)$ can be periodic. 

\textit{Center of mass of iTCP.---}
Let us numerically demonstrate the iTCP for the system with open boundaries. 
We employ the density matrix renormalization group (DMRG) method \cite{Schollwock}.
For the DMRG calculations, by employing the Jordan-Wigner transformation, we map the interacting RM model into the $S=1/2$ spin XXZ model and use the TeNPy Library \cite{Tenpy}.

\begin{figure}[t]
\centering
\includegraphics[width=8.5cm]{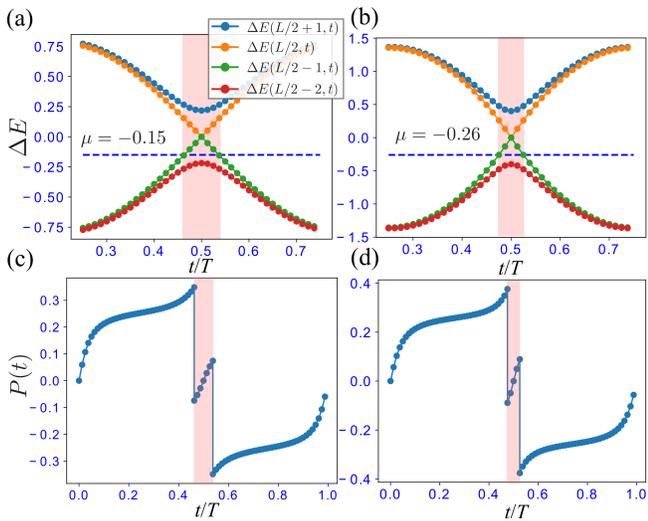}
\caption{Excitation spectrum $\Delta E$: (a) (D1)-pump protocol. (b) (D2)-pump protocol.
The behavior of the CoM with fixed $\mu$: (c) (D1)-pump protocol with $\mu=-0.15$, (d) (D2)-pump protocol with $\mu=-0.26$. 
In the red shaded area, the system includes $L/2-1$ particles, and in other regions, $L/2$-particles. For all data, $L=64$.}
\label{Fig3-2}
\end{figure}
To begin with, for the system with open boundaries and $\Delta_{j}=0$, we have calculated entanglement spectrum and determined the transition point on $J_1-V_{in}$ plane, as shown in Fig.~\ref{Fig1}(c) \cite{DMRG_PB}. The gapless phase boundary is consistent with that of the periodic case, 
determined by the quantized Berry phase $\gamma$. 

As interesting protocols, 
together with $\Delta_j(t)=(-1)^{j}\Delta_0\sin (2\pi t/T)$,
we set the following two concrete protocols. 
(D1)-pump protocol loop: $J_1=0.7$, $J_2=1$, $V_{in}(t)=1+0.5\cos(2\pi t/T)$, $\Delta_0=0.5$. 
(D2)-pump protocol loop: $J_1(t)=0.7+0.25\cos(2\pi t/T)$, $V_{in}=1$, $\Delta_0=1$.
The former is an interaction sweeping protocol and the latter is a trivial protocol at the non-interacting case, which schematic figure is shown in Fig.~\ref{Fig1} (c). 
Both protocol loops wrap the gapless phase boundary line.

\begin{figure}[t]
\centering
\includegraphics[width=8.5cm]{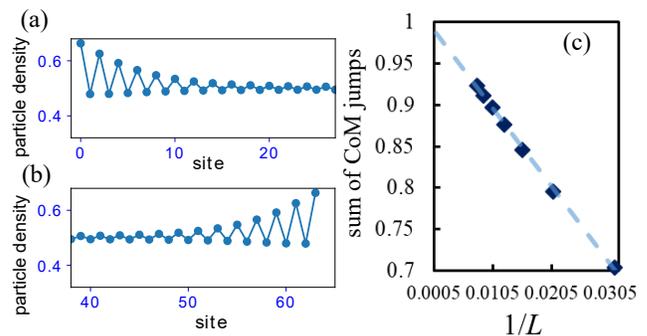}
\caption{The density distribution including left and right edge state at $t/T=0.49865$ (a) and $0.50183$ (b) in (D1)-pump protocol. 
The density distribution is defined by $\langle \Psi(L/2)|n_{j}|\Psi(L/2)\rangle$, where $|\Psi(L/2)\rangle$ is a ground-state with $L/2$ particles.
The value of CoM is $\pm 0.42287$. The system size is $L=64$.
(c) System size dependence of the total jump of the CoM with $\mu=-0.15$. 
Each jumps of the CoM are induced by changing the particle number and occurs at $t/T=0.46272$ and $0.53776$.}
\label{Fig3}
\end{figure}
The TCP for a finite system with open boundaries is characterized by the jump of the centor of mass (CoM) \cite{Hatsugai,Nakagawa,Greschner}. 
The CoM is measured under the ground canonical ensemble \cite{Hatsugai}, i.e., the system contact with a particle reservoir.
We adapt this for  the interacting system.
The CoM is given by  
\begin{eqnarray}
P(t)=\sum^{L-1}_{j=0}\langle \Psi(t)|\biggl[\frac{j-j_0}{L}\biggr]n_{j}|\Psi(t)\rangle, \nonumber
\end{eqnarray}
where $|\Psi(t)\rangle$ is a many-body ground-state at the time $t$ under ground-canonical ensemble and $j_0=(L-1)/2$.
The particle number $N_e$ of $|\Psi(t)\rangle$ is
determined by the chemical potential $\mu$ as
 $\Delta E(N_e,t) < \mu $ where $\Delta E(N_e,t)=E(N_e+1,t)-E(N_e,t)$ \cite{Chen,delE}.
(Finite chemical potential simply breaks the PH symmetry.)
Figure~\ref{Fig3-2} (a) and (b) are the numerical results of $\Delta E$ around $N_e=L/2$ for both (D1)- and (D2)-pump protocol. We find clear ingap states for a finite $V_{in}$. 
It implies the number of particles that satisfies $\Delta E(N_{e}) > \mu$ changes. 
This is gap closing as for the grand canonical hamiltonian $H_{\rm{gRM}}-\mu N$. This gap closing can be understood due to the edge states (as shown in later). It breaks the adiabaticity. It does not affect any experimental observables since the edge states are never pumped in a realistic finite speed pump.

The topological invariant $I$
of the system with boundaries is a sum of the jumps of CoM as \cite{Hatsugai},
\begin{eqnarray}
\sum_{t_i}  P(t) \Bigr|^{t_i+0}_{t_i-0}
\equiv \sum\Delta P(t_i)\xrightarrow{L\to\infty} -I. \nonumber
\end{eqnarray}
The BEC implies $I=C$ \cite{Hatsugai} which we will numerically confirm for an interacting case in later.

We set the chemical potential $\mu$ to determine the system particle number $N_e$ for each times. Figure~\ref{Fig3-2} (c) and (d) are the results of the behavior of the CoM. 
Here, the DMRG simulations find that for both (D1)- and (D2)- pump protocols, when the energy of the ingap state is equal to the chemical potential, the CoM jumps appear. 
The jump of $\Delta P(t_i)$ is associated with the change of the total number of particles. It can be attributed to the edge state (localized gapless mode). 
Actually, as displayed in Fig.~\ref{Fig3} (a) and (b), the ingap-states are left/right edge modes. 
When the total number of particles is changed, the edge states induces the change of the density distribution near the boundaries. It induces the single jump $\Delta P(t_i)$ for the CoM.

Furthermore, as shown in Fig.\ref{Fig3} (c) we observed for (D1)-pump protocol that as increasing the system size, the total sum of the jump approaches a integer value $I$:
$\sum_{t_i}\Delta P(t_i) \xrightarrow{L\to\infty} -I. $
Here $I$ is integer. 
This is the topological nature of the TCP. 
From these facts, with open boundaries and interactions, the (D1)- and (D2)-pump protocols exhibit the iTCP. 
This is due to the bulk even with boundaries although its quantization is clear by interpreting by the jump of the CoM. This is the BEC for the TCP.

\begin{figure}[t]
\centering
\includegraphics[width=8.5cm]{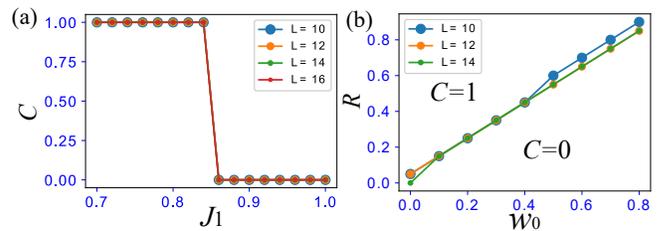}
\caption{(a) The behavior of $C_{N}$ of (D1)-pump protocol as varying $J_1$.
(b) The distribution of $C_{N}$ in (D1)-pump protocol under the perturbation $\delta V$. For all results, there is almost no system size dependence.}
\label{Fig4}
\end{figure}

\textit{Many--body Chern number and bulk--edge correspondence.---}
The CoM obtained in Fig.~\ref{Fig3-2} (c) and (d) implies the presence of the TCP in the bulk.
To verify it, 
by using ED we calculated the many-body Chern number $C$ for the periodic system. Numerically, $C$ is calculated by a discretization method \cite{Fukui-Hatsugai-Suzuki}. Focusing on (D1)-protocol, we calculated $C$ as varying the parameter $J_1$. 
The result is shown in Fig.~\ref{Fig4} (a). At $J_1=0.7$ we see $C=1$. 
(D1)-pump protocol indicates the presence of the iTCP in the bulk. Accordingly, from the result in Fig.~\ref{Fig3} (c), $C=I$ is verified, i.e., the BEC is confirmed in the iTCP. 
We have also analyze the system by the MF approximation [36] that is consistent with that of the DMRG and ED of the relative small system size. In Fig.~\ref{Fig4} (a), we also find that as increasing $J_1$, a topological phase transition occurs where $C$ suddenly changes from one to zero. This is because the pump protocol no longer wraps the gapless phase boundary (topological obstruction).

\textit{Stability of the TCP.---}
Although we start from existence of the SPT phase associated with its topological transition, iTCP is stable for any finite perturbation as far as the gap remains open. As an example of the symmetry breaking perturbation, let us introduce an on-site potential: $\delta V = \sum_{j\in even}w_0 n_j$.
Since $\delta V$ breaks the PH and BCI symmetries, at $\Delta_j(t)=0$ plane it breaks quantization of the Berry phases and the SPT is lost. 
Then, what occurs to the Chern number $C$ when the pump protocol loop is made smaller under a finite $\delta V$?  
To this end, we introduce a parameter $R$ to control the size of the protocol loop in (D1)-pump protocol: $J_1=0.7$, $J_2=1$, $V_{in}=1+R\cos(2\pi t/T)$, $\Delta_j=(-1)^{j}(R/2)\sin(2\pi t/T)$.
The phase diagram of $C$ on $w_0-R$ plane is shown in Fig.~\ref{Fig4} (b). 
As $R$ decreases, $C$ transitions at a certain point, where the protocol loop intersects a gapless phase boundary. We expect that the TCP by (D1)-pump protocol is somewhat robust against the perturbations.
This result implies that even for a finite $w_0$, the TCP is robust since a gapless phase boundary line (topological obstruction) exists within the protocol loop.

\textit{Experimental realization.---}
Our target model and pump protocol can be feasible for real experiments. 
In a coldatom optical lattice, $\Delta_j$ term is fully controllable by adjusting a double well optical lattice \cite{Nakajima,Lohse}. 
On the other hand, full control of interactions has not yet been achieved in real experimental systems. However, the implementation of the controllable interaction is feasible. 
For example, our target shape of the interaction can be implemented by selecting the kind of atom appropriately, 
such as a dipolar atom \cite{Lahaye} (e.g., Cr \cite{dePaz}, Er \cite{Baier} and Dy \cite{Lu}) and by fine-tuning spatial electric and/or magnetic external field patterns. Moreover, even if our interaction condition can be relaxed: $V_{j\in odd}=V_{out}\neq 0$ the iTCP persists \cite{SM}. This is an experimentally favorable situation. 

\textit{Conclusion.---}
Based on a simple Rice-Mele model with interactions, we have proposed the notion of the iTCP based on the topological phase transition of the SPT phase. 
A gapless phase boundary line is a topological obstruction. 
Although the SPT phases with gap closing phase transition is useful as a starting point, TCP is stable for the symmetry breaking perturbation as far as the gap along the pump is stable.
We numerically demonstrated the presence of the iTCP, and also observed that the BEC is confirmed in the interacting case. Also experimental pump protocols are proposed based on the iTCP.

\textit{Acknowledgments.---}
The work is supported in part by JSPS
KAKENHI Grant Numbers JP17H06138 (Y.K, Y.H.).

\clearpage

\renewcommand{\thesection}{S\arabic{section}} 
\renewcommand{\theequation}{S\arabic{equation}}
\renewcommand{\thefigure}{S\arabic{figure}}
\setcounter{equation}{0}
\setcounter{figure}{0}
\section*{\large{Supplemental Material}}
\section{Mean field study and its topological bands}
Let us explain the MF theory for the interacting Rice-Mele model shown in the main text. 
We will show the band spectrum to characterize the presence of the TCP by employing the MF theory.
First, let us set $J_{j\in even(odd)}=J_{1(2)}$, where $J_{1(2)}$ is a real value and $\Delta_j=0$ in $H_{\rm gRM}$ shown in the main text. 
To begin with we consider the SSH model with the interactions $V_{in}$.

To begin with we decouple the interaction term:
\begin{eqnarray}
V_{in}\sum_{j\in even} n_j n_{j+1}\to -\frac{V_{in}}{4}\sum^{(L-1)/2}_{\ell =0}\biggl[\gamma_1 {\bf f}^{\dagger}_{\ell} \hat{\sigma}_x {\bf f}_{\ell} 
+\gamma_2 {\bf f}^{\dagger}_{\ell} \hat{\sigma}_z {\bf f}_{\ell}\biggr].\nonumber\\
\label{Int_decouple}
\end{eqnarray}
Here, the constant term is dropped, $\hat{\sigma}_{x(z)}$ is the x(z)-component Pauli matrix, ${\bf f}^{\dagger}_{\ell}=(c^{\dagger}_{2\ell}, c^{\dagger}_{2\ell+1})$, $\gamma_1$ and $\gamma_2$ are mean fields describing the expectation values for the bond order within a unit cell and the charge-density wave (CDW) order, respectively, which are given by $\gamma_1=\langle c^{\dagger}_{2\ell}c_{2\ell+1}\rangle$, $\gamma_1=\langle n_{2\ell}-n_{2\ell+1}\rangle$, where $\langle \cdot \rangle$ represents an expectation value of the groundstate of the system. 
$|\gamma_1|>0$ signals the bond-order in unit cells, and $|\gamma_2|>0$ signals the CDW order.

\begin{figure}[t]
\centering
\includegraphics[width=7cm]{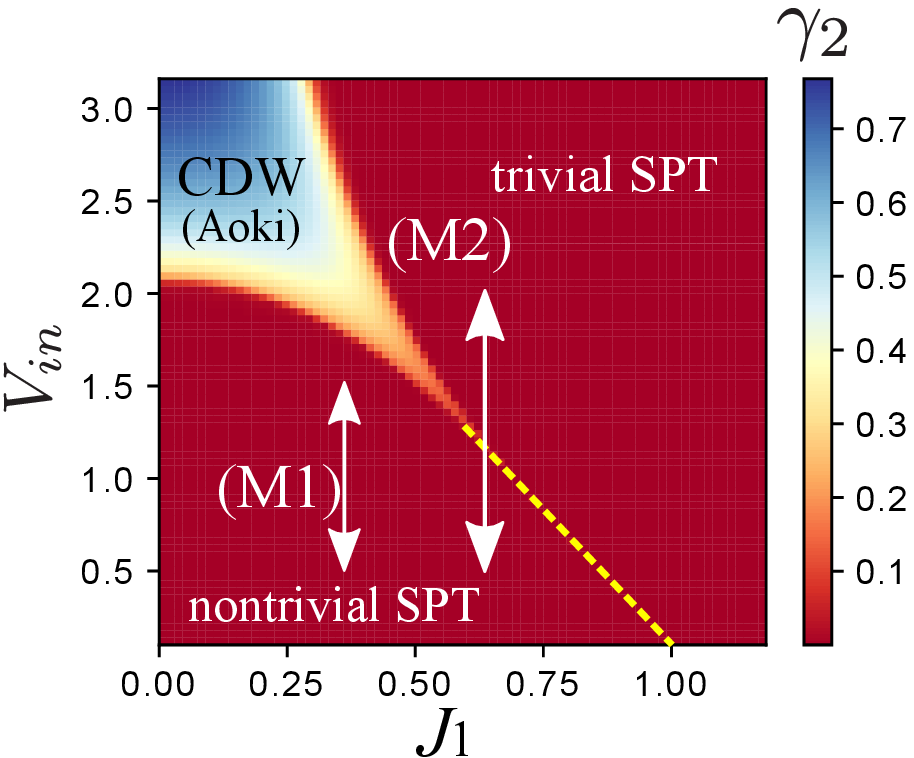}
\caption{MF phase diagram and (M1)- and (M2)-pump protocols.}  
\label{FigS1}
\end{figure}

Let us substitute Eq.~(\ref{Int_decouple}) into $H_{\rm gRM}$ of Eq.~(1) in the main text. 
For the SSH parameter case, 
the bulk momentum Hamiltonian is written as the following Bloch sphere representation: 
\begin{eqnarray}
h_{\rm gRM}(k)&=&\sum_{\alpha=x,y,z}d_{\alpha}(k)\cdot \hat{\sigma}_{\alpha}, \nonumber
\end{eqnarray}
where 
$d_x(k)=-(J_1+\Gamma_1)-J_2\cos k$,
$d_y(k)=-J_2 \sin k$, $d_z(k)=|\Gamma_2|$, $\Gamma_1=(V_{in}/2)\gamma_1$ and $\Gamma_2=(V_{in}/2)\gamma_2$.
Here, if $|\gamma_2|>0$, only in this form, 
the chiral symmetry of the SSH model looks broken since a finite $\Gamma_2$ leads to $d_z \neq 0$. 

In the MF treatment, $\gamma_1$ and $\gamma_2$ in zero-temperature limit can be directly calculated through the self-consistent way \cite{Bermudez,Kuno}. Figure~\ref{FigS1} is the global phase diagram obtained from the MF theory.
In particular, in Ref.~\cite{Bermudez}, the mean field treatment is well explained in the context of large $N$ calculations. Also, the detailed explanation of the mean field treatment for the interacting SSH model is given in the supplemental material in Ref.~\cite{Kuno4}.
\begin{figure}[b]
\centering
\includegraphics[width=8.5cm]{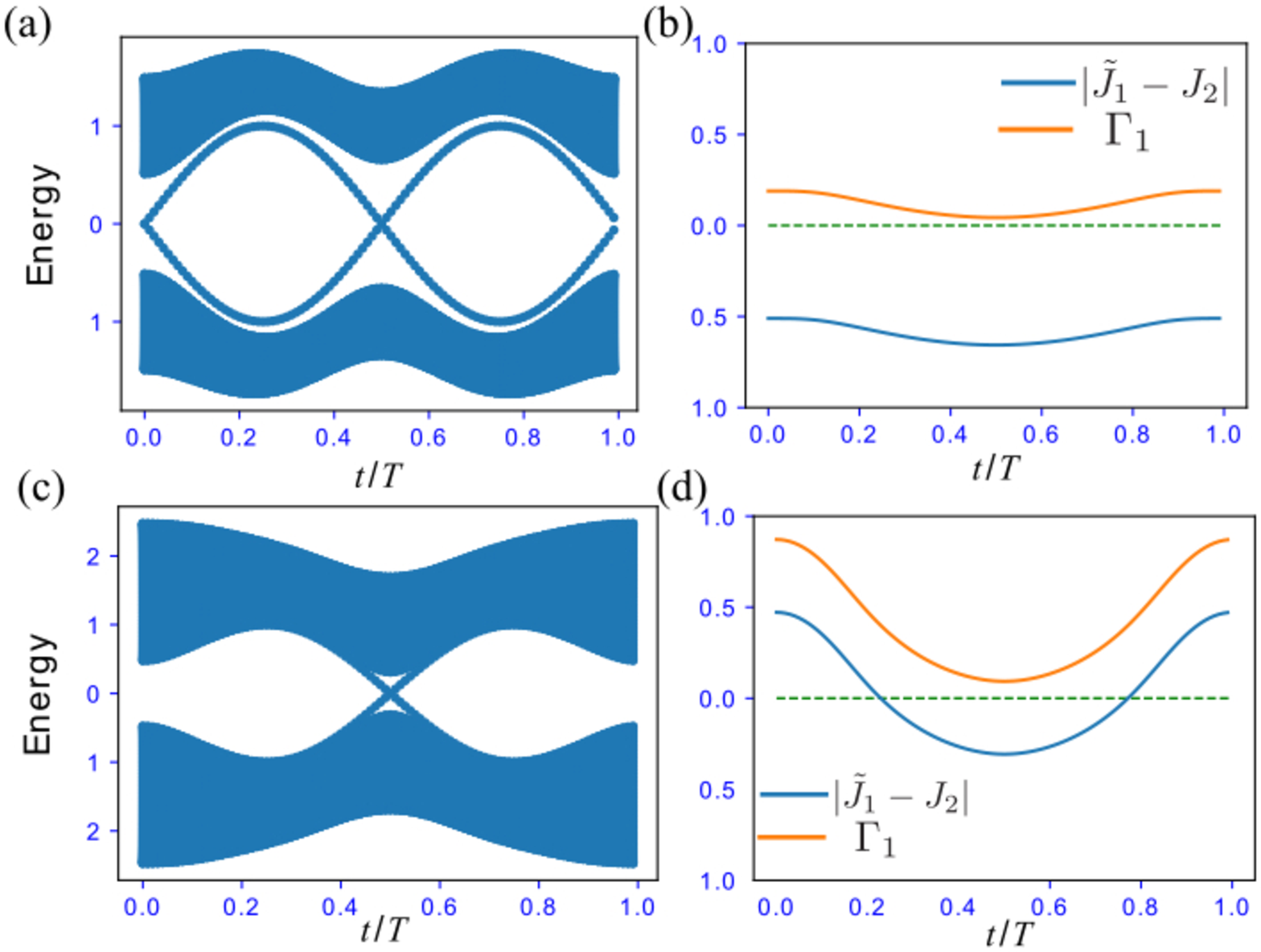}
\caption{(a) The energy spectrum for (M1)-pump protocol: $J_1=0.3$ $V_{in}=1-0.5\cos\theta$, $\Delta_0=0.5$. 
(b) The behaviors of $\tilde{J}_{1}-J_2$ and $\Gamma_1$ for (M1)-pump protocol.  
(c) The energy spectrum for (M2)-pump protocol: $J_1=0.6$ $V_{in}=1.25-0.75\cos\theta$, $\Delta_0=0.5$. 
(d) The behaviors of $\tilde{J}_{1}-J_2$ and $\Gamma_1$ for (M2)-pump protocol.}  
\label{FigS2}
\end{figure}

We investigate the presence of the iTCP in the MF level.
Let us show some concrete examples. Here, we set $\Delta_j(t)=(-1)^{j}\Delta_0 \sin(2\pi t/T)$. 
Also, to give a pump protocol loop we introduce an interaction sweeping defined by $V_{in}\to V_{in}(t)$. 
In what follows, we set $J_2=1$ and consider concrete pump protocols as shown in Fig.~\ref{FigS1}: (M1) $J_1=0.3$, $V_{in}=1+0.5\cos (2\pi t/T)$, $\Delta_0=0.5$, (M2) $J_1=0.6$, $V_{in}=1.2+0.75\cos (2\pi t/T)$ and $\Delta_0=0.5$. Here, $t$ varies from $0$ to $T$.
In (M1) and (M2)-protocols, hopping $J_1$ and $J_2$ does not change at all along time evolution.
At this time, the MF treatment can calculate the instantaneous values of MFs for each time $t$, i.e, we can obtain the values of $\Gamma_1(t)$ and $\Gamma_2(t)$. By using this values, one can directly obtain instantaneous energy spectrum including the effects of interactions under open boundary condition \cite{Klein}. One could investigate the edge mode behavior depending on $t$.
Figure~\ref{FigS2} (a) and (c) are the result of the energy spectrum for (M1) and (M2) pump protocols. 
Interestingly enough, we find clear signature of the interacting effect of the edge modes.  
For (M1)-protocol result, the left and right edge mode appears even for finite interaction. However since (M1)- protocol does not wrap the phase boundary line nor connect to two different phases, the edge modes appear and cross even at $t/T=0$,$1/2$, where SSH model is recovered. 
This means the (M1)-protocol does not exhibit the iTCP.
According to the bulk--edge correspondence (BEC) of the TCP \cite{Hatsugai}, we expect that the M1 protocol does not exhibit the TCP in the bulk. Also in Fig.~\ref{FigS2} (b), we plot $\Gamma_{1}$ and the difference between the effective coupling $\tilde{J}_1\equiv J_1+\Gamma_1$ and $J_2$, $\tilde{J}_{1}-J_2$. For all time , $\tilde{J}_1> J_2$. 
On the other hand, as shown in Fig.~\ref{FigS2} (b), (M2)-protocol exhibits the single crossing of the left and right edge mode at $t/T=1/2$, but interestingly at $t=0$ point, the spectrum is gapped out, no crossing of the edge modes. Figure \ref{FigS2} (d) shows the behaviors of $\tilde{J}_{1}-J_2$ and $\Gamma_1$ along the pump protocol. Here, due to the interaction-induced bond order $\Gamma_1$, $J_1$ is somewhat corrected and leads to connect the different phases without gap closing. 
Theses results implies that (M2)-protocol possesses the iTCP.

In the MF level, we clarified the presence of the iTCP. In particular, an interaction sweeping iTCP exists.

\section{Effects of inter-unit cell interactions}

\begin{figure}[h]
\centering
\includegraphics[width=6cm]{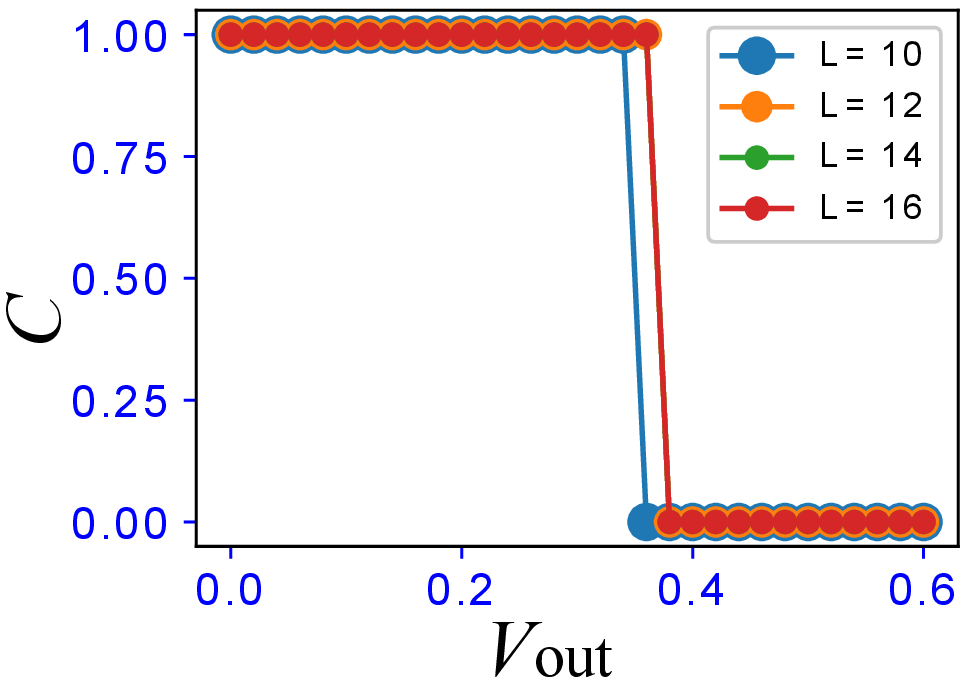}
\caption{The behavior of $C_{N}$ of (D1)-protocol as varying $V_{out}$, $J_1=0.7$.
(D1)-protocol: $V_{in}=1-0.5\cos\theta$, $\Delta_0=0.5$.
$L/2$-particle system. For all results, there is almost no system size dependence.}
\label{FigS3}
\end{figure}

It may be difficult for real experiments such as coldatoms \cite{Lohse, Nakajima} to implement only $V_{in}$ interactions. 
Here, we investigate the effects of inter-unit cell interaction: $V_{j\in odd}=V_{out}\neq 0$.

For the periodic boundary case, we calculated the many-body Chern number $C$. 
See Fig.~\ref{FigS3}, we find that as increasing $V_{out}$ for (D1)-pump protocol defined in the main text, 
the bulk iTCP is robust up to $V_{out}/V_{in}\sim 0.38$. 
That is, even for a finite $V_{out}$, the iTCP exists. 

\begin{figure}[b]
\centering
\includegraphics[width=8.5cm]{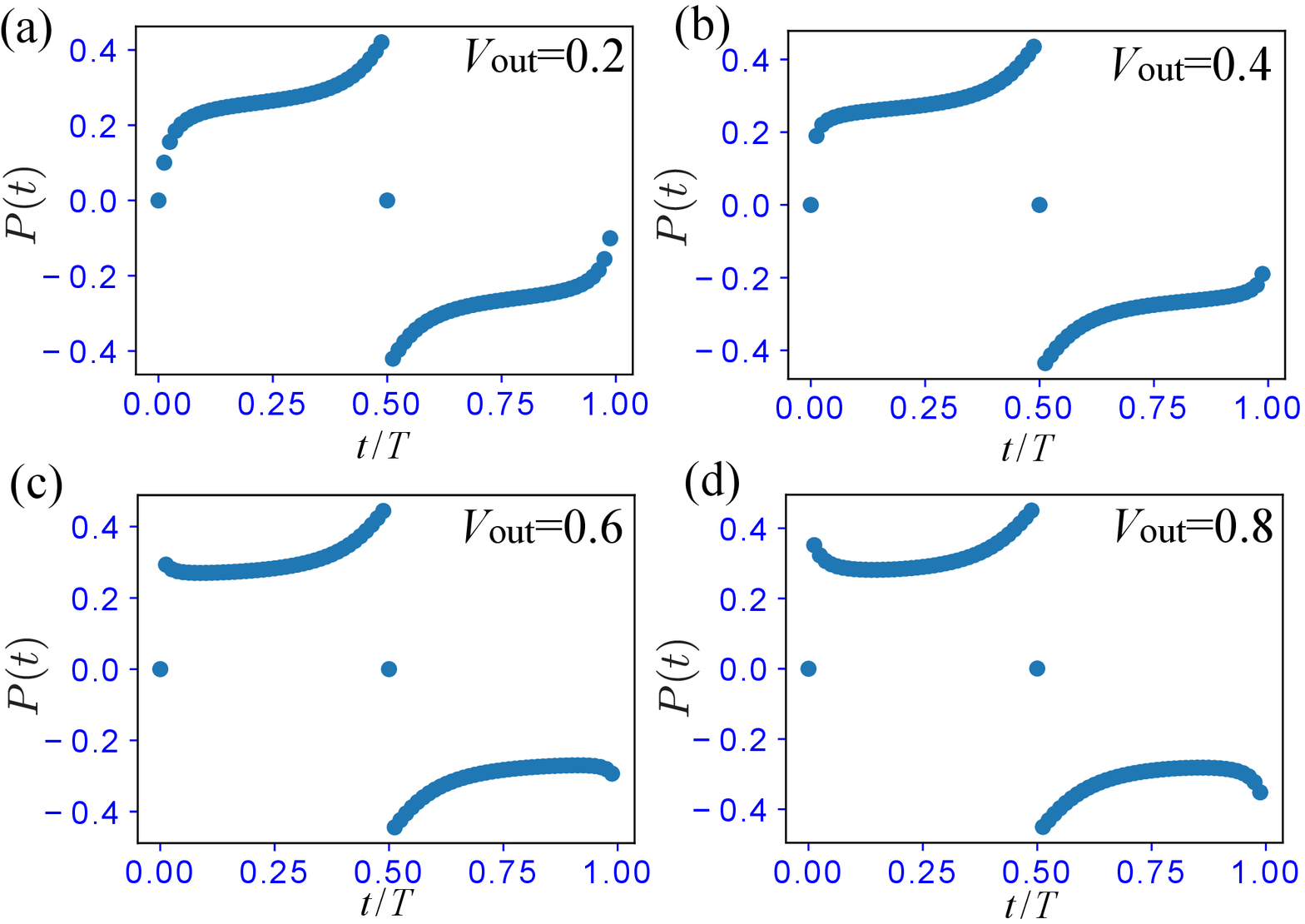}
\caption{The behavior of CoM under inter unit cell interaction $V_{out}$.
(a) $V_{out}=0.2$,
(b) $V_{out}=0.4$,
(c) $V_{out}=0.6$,
(d) $V_{out}=0.8$. The system size is $L=64$.}  
\label{FigS4}
\end{figure}

We also show the behavior of the CoM under effects of inter-site interaction $V_{out}$. For a finite system with open boundary condition, we calculate the CoM by using the DMRG. We employ the (D1)-pump protocol as shown in the main text. We calculated the CoM on the case that the number of particles was fixed to $L/2$. Then, the TCP is expected to be characterized to a single jump at $t/T=1/2$ if the TCP presents \cite{Nakagawa, Hatsugai,Greschner}. This jump occurs by exchanging the occupancy of the left and right edge modes.
Here, the value of the jump of the CoM is strictly one in large system size. Figure~\ref{FigS4} is the results of the behavior of the CoM under effects of inter-site interaction $V_{out}$. 
For $V_{out}=0.2$, the jump of the CoM clearly appears around $t/T=1/2$. Then for $V_{out}=0.4$, we observes a small jump between $t/T=0$ and $1$, which is a signal of the breakdown of the TCP. As shown in Fig.~\ref{FigS4} (c) and (d), we further increase $V_{out}$, the jump becomes larger, and the TCP completely breaks. From these results, the breakdown threshold of the TCP is around $V_{out}\sim 0.4$, close to the phase transition point of the many-body Chern number $C$ in Fig.~\ref{FigS3} in the main text. 
For the finite system with open boundary condition, the TCP characterized by the CoM is robust up to $V_{out}/V_{in}\sim 0.4$.

Such a finite interaction combination of $V_{in}$ and $V_{out}$ could be feasible for future optical lattice systems with dipole-dipole interactions \cite{dePaz,Baier,Lu}. 

\begin{thebibliography}{99}
\bibitem{SPT_dis}
The situation can be somewhat relaxed. 
In the case without SPT phase with nontrivial $\gamma$, 
if edge modes appear, the iTCP can be constituted.
At least the existence of SPT phase is the basis for generating the iTCP.
\bibitem{BCS}
The bond-centered inversion transformation is defined by $c_j\to c_{L-j-1}$.
\bibitem{SPT_shift}
From the MF theory, 
the shift of the SPT phase boundary line from $J_1 = 1$ is attributed to the enhancement of a bond order induced by $V_{in}$.
See \cite{SM}.
\bibitem{DMRG_PB}
The phase boundary was determined by the behavior of the entanglement entropy and the behavior of the center of mass under very small inversion breaking potential.
\bibitem{delE}
Consider an energy of many-body state with particle $N_e$, $E_{N_e}$. Then $N_e$ of the grand canonical ground state needs to satisfy $E_{N_e+1}-\mu (N_e+1)>0$ and
$E_{N_e}-\mu N_e <0$. It implies $\Delta E_(N_e,t)< \mu $.
\end{thebibliography}
\end{document}